\documentclass[11pt]{article}
\newcommand{\newsection}{\setcounter{equation}{0}
 \section}

\def\tfr#1#2{{\textstyle{#1\over#2}}}
\def\~#1{{\bf\widetilde{\mit#1}}}
\def\ds{\displaystyle}
\def\d{{\rm d}}
\def\l{\ell}
\def\p{Painlev\'e}
\def\bk{B\"acklund}

\def\bts{B\"ack\-lund transformations}
\def\peq{\p\ equation}
\def\peqs{\p\ equations}

\begin{document}
\setcounter{equation}{0} \setcounter{section}{0}
\title{\Large \bf Discrete equations and the singular manifold method}
\author{{\bf P.G.\ Est\'evez$^1$ and P.A.\ Clarkson$^2$}\\
$^1$Area de Fisica Te\'orica, Facultad de Ciencias\\ Universidad de
Salamanca\\ 37008
Salamanca, Spain\\
$^2$Institute of Mathematics and Statistics \\
University of Kent at Canterbury \\ Canterbury, Kent, CT2 7NF, U.K.}
\maketitle
\begin{abstract}

The Painlev\'e expansion for the second \p\ equation (PII) and fourth \p\
equation (PIV) have two branches. The singular manifold method therefore
requires two singular manifolds. The double singular manifold method is used
to derive Miura transformations from PII
and PIV to modified \p\ type equations for which auto-\bk\ transformations
are obtained.
These auto-\bk\ transformations can be used to obtain discrete equations.
\end{abstract}

\newsection{Introduction}
An specially important subject related to the study of discrete equations
is that
concerning the discretization of \p\ equations (cf.\
\cite{BC94,GR98,NSKGR96,RG92}).
Recently there has been substantial interest in the discrete \p\ equations
(dP1--dPVI), which in
a variety of physical applications and indeed dPI and dPII were first
discovered in physical
situations \cite{refBK,refGM,refPS}. In the continuous limit, the discrete
\p\ equations yield
their corresponding continuous \p\ equation. This is not all as they have a
variety of other
properties in common with their continuous counterparts. For example have
Lax pairs, bilinear
representations, \bk\ transformations and particular solutions for certain
parameter values,
expressible in terms of rational functions or discrete special functions
(see, for example,
\cite{refGNR} and the references therein).

The discrete equations, though, are much richer than the continuous
equations. They have a
host of properties which are lost in the continuous limit. Perhaps the most
fundamental
difference is that the continuous \p\ equations have a unique canonical
form whereas
this is not true for the discrete equations: for any discrete \p\ equation
there
exists a number of different forms.

It is well known that continuous \p\ equations satisfy the \p\ property (cf.\
\cite{Davis,Ince}) which means that all their solutions $y(x)$ can be
locally expressed as
\begin{equation}
y(x)=\sum_{j=0}^{\infty}a_j(x-x_0)^{j-\alpha},\label{1.1}
\end{equation}
in a neighborhood of the movable singularity $\phi\equiv x-x_0=0$, where
$x_0$ is an arbitrary
constant, $\alpha$ is a positive integer and $a_j$, $j=0,1,\dots$, are
constants to be determined.
More precisely, the singular manifold method (SMM)
\cite{Weiss83} consists of truncation of the series (\ref{1.1}) to the
constant level
$$
y(x)=\sum_{j=0}^{\alpha}a_j(x-x_0)^{j-\alpha},
$$
This condition seems to be very restrictive, nevertheless it provides much
information about a
given equation (cf.\ \cite{CJP98,Weiss83}).

In the present paper we shall apply the SMM to the second and fourth \p\
equations, namely PII and PIV, which are usually written as
\begin{eqnarray}
&{\rm PII}:\qquad y''&=2y^3+xy+\l,\label{1.2}\\
&{\rm PIV}:\qquad y''&={(y')^2\over 2y}+\frac32
y^3+4xy^2+2(x^2-2\l)y+{\mu\over y},\label{1.3}
\end{eqnarray}
where $'\equiv\d/\d x$ and $\l$ and $\mu$ are constants (cf.\ \cite{Ince}).

\subsection{Two \p\ branches and the singular manifold method}
These equations (\ref{1.2},\ref{1.3}) have the property that the leading
analysis of the expansion
(\ref{1.1}) provides $\alpha=1$ and $a_0=\pm 1$, where the $\pm$ sign in
$a_0$ indicates that
there are two possible \p\ branches.

Equations with two \p\ branches and how to extend the SMM to them have been
studied in previous papers (cf.\ \cite{EG93,EG97}). According to these,
when an equation,
such as (\ref{1.2}) or (\ref{1.3}), has two \p\ branches, then the
truncation of (\ref{1.1})
should be made for both branches simultaneously. This means that we should
work with truncated
solutions $\~y$ of the following form \cite{EG97}
\begin{equation} \~y=y+{g'\over g}-{h'\over h},\label{1.4}\end{equation}
where $y(x)$ is also a solution of the equation, and $g(x) =0$ is the
singularity for the $+$
expansion and $h(x)=0$ the singularity for the $-$ expansion. Henceforth we
shall call $g$ and
$h$ {\it singular manifolds} \cite{Weiss83}.

The SMM usually requires that each coefficient in the different powers of
the singular manifold
that arises from substitution of the truncated expansion in the
differential equation be equal to
zero. Due to the non-linearity of PII and PIV, the substitution of
(\ref{1.4}) in (\ref{1.2}) and
(\ref{1.3}) respectively provides terms that mix the powers in $h$ and $g$.
To solve this problem
we use a {\it decoupling ansatz} \cite{EG97} given by
\begin{equation} {g'\over g}{h'\over h}=A{g'\over g}+B{h'\over
h},\label{1.5}\end{equation}
where $A$ and $B$ are functions of $y$, and $g$ and $h$ to be determined
from the truncation
itself. Furthermore, taking the derivative of (\ref{1.5}) with respect to
$x$ we have
\begin{equation}
A'=A(r-A-B)\qquad \qquad B'=B(v-A-B),\label{1.6}\end{equation}
where $v={g''/g'}$ and $r={h''/h'}$.

Within this framework, the objective of this paper is to prove that for PII
and PIV the improved
version of the SMM including two singular manifolds provides the following
results:

\begin{itemize}
\item[(i)] Modified versions of PII and PIV (namely mPII and mPIV).

\item[(ii)] Miura transformations between PII and mPII and between PIV and
mPIV.

\item[(iii)] \bk\ transformations for mPII and mPIV.

\item[(iv)] By regarding the parameter $\l$ as a discrete variable,
discrete equations
can be derived associated with mPII and mPIV.

\item[(v)] Solutions of these discrete equations allow us to construct
solutions of PII and
PIV as a ``linear" superposition of solutions of the discrete equations.
\end{itemize}

\newsection{\p\ II}
Substituting (\ref{1.4}) into PII (\ref{1.2}) and the use of (\ref{1.5}) to
decouple the
crossed terms yields
\begin{eqnarray}
A&=&y+\tfr{1}{2}v,\label{2.2}\\
B&=&-y+\tfr{1}{2}r,\label{2.3}\\
0&=&v'+v^2-6y^2-x+6A(2y+B-A),\label{2.4}\\
0&=&r'+r^2-6y^2-x-6B(2y+B-A).\label{2.5}\end{eqnarray}
Using (\ref{1.6}) in (\ref{2.2}--\ref{2.3}) gives
$$AB=k, $$ 
where $k$ is a constant. By substituting (\ref{2.2}--\ref{2.3}) into
(\ref{2.4}--\ref{2.5}), the
following singular manifold equations are obtained
\begin{equation} v'-\tfr12{v^2}+6k-x=0,\qquad\qquad
r'-\tfr12{r^2}+6k-x=0.\label{2.7}\end{equation}

\subsection{A Miura transformation for modified PII}
If we return to equation (\ref{1.6}), substituting $A$ and $B$ as given by
(\ref{2.2}) and
(\ref{2.3}), respectively, and using (\ref{2.7}), we have
\begin{equation}
y'-y^2-\tfr12{x}+v'+2k=0,\qquad-y'-y^2-\tfr12{x}+r'+2k=0,\label{2.8}\end{equation}
which means that the expression $y'-y^2-\tfr12{x}$ depends only on the
singular manifold $g$
whilst $-y'-y^2-\tfr12{x}$ depends only on $h$. Hence it is useful to
define the following
functions $m'$ and $n'$ as follows
\begin{equation} 2m'=y'-y^2-\tfr12{x},\qquad
2n'=-y'-y^2-\tfr12{x}.\label{2.9}\end{equation}
With the aid of (\ref{1.2}), these equations (\ref{2.9}) can be integrated
to give
\begin{equation} 2m=(y')^2-(y^2+\tfr12{x})^2-(2\l-1)y,\qquad
2n=(y')^2-(y^2+\tfr12{x})^2-(2\l+1)y,\label{2.10}\end{equation}
respectively. In order to identify the equations that $m$ and $n$ satisfy,
we take the derivative
of (\ref{2.9}), which gives
\begin{equation} 2m''=-4ym'+\left(\l-{\tfr12}\right),\qquad
2n''=4yn'-\left(\l-{\tfr12}\right).\label{2.11}\end{equation}
Now if we take $y'$ from (\ref{2.9}) and  $y$ from (\ref{2.11}) and
substitute them in
(\ref{2.10}), then we obtain the following equation for $m$ and $n$
\begin{equation} {(M_{\l}'')^2\over M_{\l}'}+4(M_{\l}')^2
+2(xM_{\l}'-M_{\l})-{(2\l-1)^2\over16M_{\l}'}=0,\label{2.12}\end{equation}
where $m=M_{\l}$ and $n=M_{\l+1}$. It is easy to prove that this equation
has the \p\ property. In
fact, equation (\ref{2.12}) is the potential version of the equation 34 of the
Gambier classification  \cite{G10} (see also \cite{FA82,Ince,RG92}), which
is commonly
referred to as P34. Indeed, if we set $M_{\l}'=Q_{\l}$, then (\ref{2.12})
becomes
\begin{equation} Q_{\l}''-{(Q_{\l}')^2\over
2Q_{\l}}+4Q_{\l}^2+xQ_{\l}+{(\l-{\tfr12})^2\over
8Q_{\l}}=0,\label{2.15}\end{equation}
which is precisely P34. Equations (\ref{2.9}) and (\ref{2.11}) can be
written as
$$\begin{array}{ll}
\ds 2Q_{\l}=y'-y^2-\tfr12{x},\qquad\qquad &\ds
2Q_{\l+1}=-y'-y^2-\tfr12{x},\\[10pt]
\ds y={-2Q_{\l}'+\l-{\tfr12}\over
4Q_{\l}},\qquad\qquad &\ds y={-2Q_{\l+1}'+\l+{\tfr12}\over
4Q_{\l+1}}.\end{array}$$
These can be interpreted as a Miura transformation between PII (\ref{1.2})
and P34 (\ref{2.15})
\cite{FA82}. According to Ramani and Grammaticos \cite{RG92}, P34
(\ref{2.15}) may be also be
thought of as a modified PII (mPII), since the relationship between their
solutions is analogous
to that between solutions of the Korteweg-de Vries and modified Korteweg-de
Vries equations (see
also \cite{FA82}).

\subsection{Auto-\bk\ transformations for modified PII}
By subtracting the two equations of (\ref{2.10}), we have
\begin{equation}y(x)=M_{\l}(x)-M_{\l+1}(x),\label{2.18}\end{equation}
which combined with (\ref{2.11}) gives
\begin{equation} M_{\l+1}= M_{\l}+{4M_{\l}''-(2\l-1)\over 8M_{\l}'},\qquad
M_{\l}=M_{\l+1}+{4M_{\l+1}''+(2\l+1)\over 8M_{\l+1}'},\label{2.19}
\end{equation}
which are auto-\bk\ transformations for potential mPII.

\subsection{Linear superposition for PII}
Suppose we have two solutions $M_{\l}$ and $M_{\l+1}$ of potential mP34
(\ref{2.12}), related by
the \bk\ transformations (\ref{2.19}), then we can construct a solution of
PII (\ref{1.2}) by
using (\ref{2.18}).

\subsection{Discrete equations for potential P34}
If we let $\l\rightarrow \l-1$ in the second of equations (\ref{2.19}) then
\begin{equation} M_{\l+1}= M_{\l}+{4M_{\l}''-(2\l-1)\over 8M_{\l}'},\qquad
M_{\l-1}=
M_{\l}+{4M_{\l}''+(2\l-1)\over 8M_{\l}'}.\label{2.20}\end{equation}
Adding and subtracting these two equations gives
\begin{equation}M_{\l}'=-\,{2\l-1\over 4(M_{\l+1}-M_{\l-1})},\qquad
M_{\l}''=-\,{(2\l-1)(M_{\l+1}+M_{\l-1}-2M_{\l})\over
4(M_{\l+1}-M_{\l-1})}.\label{2.21}\end{equation} By substituting
(\ref{2.21}) in (\ref{2.12}),
the result is the nonautonomous discrete equation
\begin{eqnarray*}
&&(M_{\l-1}-M_{\l+1})\left\{\left(\l-{\tfr12}\right)\left[(M_{\l-1}-M_{\l})(M_{\
l+1}-M_{\l})+\tfr12{x}\right]\right.\\
&&\qquad\left.
-M_{\l}(M_{\l-1}-M_{\l+1})\right\}
+\left(\l-{\tfr12}\right)^2=0,
\end{eqnarray*}
where the parameter $\l$ can be interpreted as the discrete variable.

\newsection{\p\ IV}
In an analogous way as we studied PII in the previous section, in this
section we study PIV
(\ref{1.3}) by using the truncated expansion (\ref{1.4}) together with the
decoupling ansatz
(\ref{1.5}). The result is
\begin{eqnarray}
A&=&y+x+\tfr{1}{2}v,\label{3.2}\\
B&=&-y-x+\tfr{1}{2}r,\label{3.3}\\
0&=&(24x+6v)(A-y)-4x^2+8\l+v^2+2(v'+y')\nonumber\\
&&\qquad-18y^2-16A^2+36yA+8AB-2rA,\label{3.4}\\
0&=&(-24x+6r)(B+y)-4x^2+8\l+r^2+2(r'-y')\nonumber\\
&&\qquad-18y^2-16B^2-36yB+8AB-2vB.\label{3.5}
\end{eqnarray}
Equations (\ref{3.2}--\ref{3.3}) combined with (\ref{1.6}) gives $AB=k$,
with $k$ a constant. Then
substituting (\ref{3.2}--\ref{3.3}) into (\ref{3.4}--\ref{3.5}),  for the
singular manifolds we
have the equations
\begin{equation} v'-\tfr12{v^2}+6k+2x^2+8\l-2=0,\qquad
r'-\tfr12{r^2}+6k+2x^2+8\l+2=0.\label{3.7}\end{equation}

\subsection{A Miura transformation for modified PIV}
The substitution of (\ref{2.2}-\ref{2.3}) into (\ref{1.6}) gives
\begin{equation}
 y'-y^2-2xy+v'+2(k+2\l)=0,\qquad
-y'-y^2-2xy+r'+2(k+2\l)=0.\label{3.8}\end{equation}
It is reasonable therefore to define new functions $m$ and $n $ in the
following form
\begin{equation}
 2m'=y'-y^2-2xy,\qquad 2n'=-y'-y^2-2xy.\label{3.9}\end{equation}
These equations can be integrated, taking (\ref{1.3}) into account, as
\begin{equation} \begin{array}{l}
\ds 2m={{(y')^2+2\mu -(y^2+2xy)^2}\over 4y}+(2\l+1)y,\\[10pt]
\ds 2n={{(y')^2+2\mu -(y^2+2xy)^2}\over 4y}+(2\l-1)y. \end{array}
\label{3.10}\end{equation}
Furthermore, the derivation of (\ref{3.9}) provides
\begin{equation} \begin{array}{l}
\ds 2m''={2(m')^2+\mu \over y}-2(m'+2\l+1)y,\\[10pt] \ds 2n''=-{2(n')^2+\mu
\over y}+2(n'+2\l-1)y.
\end{array}\label{3.11}\end{equation}
Taking $y'$ as defined by (\ref{3.9}) and substituting in (\ref{3.10}) we have
\begin{equation} \begin{array}{l}
\ds 4(m-xm')={2(m')^2+\mu \over y}+2(m'+2\l+1)y,\\[10pt] \ds
4(n-xn')={2(n')^2+\mu \over y}+2(n'+2\l-1)y.
\end{array}\label{3.12}\end{equation}
Adding and subtracting (\ref{3.11}) and (\ref{3.12}) the result is
\begin{equation} \begin{array}{ll}
\ds y={2(m-xm')-m''\over 2(m'+2\l+1)},\qquad &
\ds{1\over y}={2(m-xm')+m''\over 2(m')^2+\mu },\\[10pt]
\ds y={2(n-xn')+n''\over 2(n'+2\l-1)},\qquad & \ds
{1\over y}={2(n-xn')-n''\over 2(n')^2+\mu
}.\end{array}\label{3.13}\end{equation}
We can eliminated $y$ by multiplication of the left equation by the right
one. Thus we then have
\begin{equation}
\left(M_{\l}''\right)^2-4\left(M_{\l}-xM_{\l}'\right)^2
+2\left(M_{\l}'+2\l+1\right)\left[2\left(M_{\l}'\right)^2+\mu
\right]=0,\label{3.14}\end{equation}
where
$m=M_{\l}$ and $n=M_{\l-1}$. Equation (\ref{3.14}) is of
\p\ type, and it can be considered as the modified version of PIV (mPIV). The
Miura transformation that relates PII and mPII are (\ref{3.11}), which can
now be written as
\begin{equation}
2M_{\l}'=y'-y^2-2xy,\qquad 2M_{\l-1}'=-y'-y^2-2xy.\label{3.15}\end{equation}

\subsection{Auto-\bk\ transformations}
Subtracting the two equations of (\ref{3.10}), and with the aid of
(\ref{3.15}), we have
\begin{equation} y=M_{\l}-M_{\l-1},\label{3.17}\end{equation}
then combined with the left part of (\ref{3.13}) yields to
\begin{equation} \begin{array}{l}
\ds M_{\l-1}=M_{\l}+{M_{\l}''-2(M_{\l}-xM_{\l}')\over
2(M_{\l}'+2\l+1)},\\[10pt]
\ds M_{\l}=M_{\l-1}+{M_{\l-1}''+2(M_{\l-1}-xM_{\l-1}')\over
2(M_{\l-1}'+2\l-1)},
\end{array}\label{3.18}\end{equation}
which are auto-\bk\ transformation for mPIV (\ref{3.14}).

\subsection{Linear superposition for PIV}
If we have two solutions $M_{\l}$ and $M_{\l+1}$ of potential mPIV
(\ref{3.14}),
related by the \bk\ transformations (\ref{3.18}), then we can construct a
solution of PIV
(\ref{1.3}) by using (\ref{3.17}).

\subsection{Discrete equations for mPIV}
As we did in the previous section, the \bk\ transformations (\ref{3.18})
can be written as
\begin{equation} \begin{array}{l}
\ds M_{\l-1}=M_{\l}+{M_{\l}''-2(M_{\l}-xM_{\l}')\over
2(M_{\l}'+2\l+1)},\\[10pt]
\ds M_{\l+1}=M_{\l} +{M_{\l}''+2(M_{\l}-xM_{\l}')\over 2(M_{\l}'+2\l+1)}.
\end{array}\label{3.19}\end{equation}
Then by addition and subtraction, it is easy to show that
\begin{equation} \begin{array}{l} \ds
M_{\l}'={2M_{\l}-(2\l+1)(M_{\l+1}-M_{\l-1})\over
 M_{\l+1}-M_{\l-1}+2x},\\[10pt] \ds
M_{\l}''={(M_{\l}'+2\l+1)(M_{\l+1}+M_{\l-1}-2M_{\l})\over
M_{\l+1}-M_{\l-1}+2x},\end{array}\label{3.20}\end{equation}
whose substitution in (\ref{3.15}) yields the discrete equation
\begin{eqnarray*}
&&(M_{\l+1}-M_{\l-1}+2x)(M_{\l}-M_{\l-1})(M_{\l}-M_{\l+1})\left[M_{\l}+2\left(\l
+{\tfr12}\right)x\right]\\
&&\qquad+\left\{2\left[M_{\l}-\left(\l+{\tfr12}\right)\right]
(M_{\l+1}-M_{\l-1})\right\}^2
+\tfr12{x}\left(M_{\l+1}-M_{\l-1}+2x\right)^2=0.
\end{eqnarray*} 
where $\l$ is the discrete parameter.

\newsection{Conclusions}
In this paper we have derived Miura transformations, modified equations and
associated discrete
equations for the second and fourth \p\ equations. Recently there have been
several studies of
the derivation of discrete equations, and in particular the discrete \p\
equations, from \bk\
transformations of the (continuous) \p\ equations
\cite{refFGR,refGNR,refGra,GR98,refGTii,refGTiii,NSKGR96,refTsegb,refTsegc}.

Hierarchies of solutions of PII and PIV are well-known (cf.\
\cite{refAirault,BCH95,PAC90,refGromaki,Gromak,refLuka,Murata,refOka,refUW}). 
Since there is an explicit relationship between PII and PIV and discrete
equations, then these Hierarchies of
solutions of PII and PIV also satisfy difference equations in addition to
the ordinary
differential equations. This is analogous to the situation for the
classical special functions,
such as Bessel, hypergeometric, Legendre, Weber-Hermite and Whittaker
functions, which satisfy
both an ordinary differential equation and a recurrence relation, which is
a discrete equation.
This provides further evidence that the \p\ equations may be thought of as
nonlinear special
functions and that there is a deep relationship between the classical
special functions, the \p\
equations and the discrete \p\ equations (see, for example,
\cite{refTRGK}).

\def\fit{\frenchspacing\it}

\end{document}